\documentclass[aps,prl,preprint,floatfix,showpacs]{revtex4-1}

\usepackage{bm}
\usepackage{epsfig}
\usepackage{comment}
\usepackage{epstopdf}
\usepackage{graphicx}
\usepackage{graphics}
\DeclareGraphicsExtensions{.eps}
\usepackage{color}
\usepackage{bigints}
\usepackage{multirow}

\usepackage{amsmath}
\usepackage{amsfonts}
\usepackage{amssymb}

\begin{document}

\title{Microscopic magnetic properties of the ferromagnetic superconductor UCoGe reviewed by X-ray magnetic circular dichroism}
\author{M.~Taupin$^{1}$}
\thanks{Present address: Low Temperature Laboratory, Aalto University, P.O. Box 13500, FI-00076 Aalto, Finland}
\author{J.-P.~Sanchez$^1$}
\author{J.-P.~Brison$^1$}
\email[]{jean-pascal.brison@cea.fr}
\author{D.~Aoki$^{1,2}$}
\author{G.~Lapertot$^{1}$}
\author{F.~Wilhelm$^3$}
\author{A.~Rogalev$^3$}
\affiliation{$^1$Univ. Grenoble Alpes, INAC-SPSMS, F-38000 Grenoble, France\\
$^{\ }$CEA, INAC-SPSMS, F-38000 Grenoble, France\\
$^2$Institute for Materials Research, Tohoku University, Oarai, Ibaraki 311-1313, Japan\\
$^3$European Synchrotron Radiation Facility (ESRF), B.P. 220, FR-38043 Grenoble, France}

\date{\today}

\begin{abstract}
The ferromagnetic superconductor UCoGe has been investigated by high field X-ray magnetic circular dichroism (XMCD) at the U-M$_{4,5}$ and Co/Ge-K edges. The analysis of the branching ratio and XMCD at the U-M$_{4,5}$ edges reveals that the U-5$f$ electrons count is close to 3. The orbital ($\sim0.70\,\mu_B$) and spin ($\sim-0.30\,\mu_B$) moments of U at 2.1\,K and 17\,T (\textbf{H}//\textbf{c}) have been determined. Their ratio ($\sim-2.3$) suggests a significant delocalization of the 5$f$ electron states. The similar field dependences of the local U/Co and the macroscopic magnetization indicate that the Co moment is induced by the U moment. The XMCD at the Co/Ge-K edges reveal the presence of small Co-4$p$ and Ge-4$p$ orbital moments parallel to the macroscopic magnetization. In addition, the Co-3$d$ moment is estimated to be at most of the order of 0.1\,$\mu_B$ at 17\,T. Our results rule out the possibility of an unusual polarisability of the U and Co moments as well as their antiparallel coupling. We conclude that the magnetism which mediates the superconductivity in UCoGe is driven by U.
\end{abstract}


\maketitle

The true coexistence of ferromagnetism (FM) and unconventional superconductivity (SC) was first discovered in UGe$_2$ under pressure \cite{Saxena2000} and then in URhGe \cite{Aoki2001} and UCoGe \cite{Huy2007}. The two last isostructural strongly correlated electron systems have the peculiarity to be SC at ambient pressure thus prone to more detailed experimental studies. UCoGe is a weak FM with a Curie temperature ($T_{Curie}$ = 2.8\,K) higher than the SC temperature ($T_{SC}$ = 0.5\,K) and an ordered moment of about 0.07\,$\mu_B$/f.u. UCoGe crystallizes in the orthorhombic TiNiSi-type structure with an Ising-like anisotropy of the magnetization, \textbf{c} being the easy-axis, \textbf{a} the hard axis and \textbf{b} the intermediate-one. A striking point is the large upper critical field $H_{c2}$ exceeding the Pauli paramagnetic limit when the magnetic field is applied along the hard magnetization axes (\textbf{a} and \textbf{b}-axis) \cite{Aoki2009}. Moreover when the field is applied very precisely along the \textbf{b}-axis an unusual inverse ``S''-shaped $H_{c2}$ curve is observed \cite{Aoki2009}. These phenomena seem closely related to the ferromagnetic instability as $T_{Curie}$ is reduced when \textbf{H}//\textbf{b} and collapses at the enhanced superconducting phase. The Sommerfeld specific heat coefficient $\gamma$, and the T$^2$ term resistivity coefficient $A$, both related to the effective mass of conduction electrons, reach a maximum value around 14\,T \cite{Aoki2011a}. This all suggests a spin-triplet type pairing mechanism untimely related to critical spin fluctuations connected to a magnetic instability.

In URhGe, neutron diffraction experiments could demonstrate that maximum $T_{SC}$ of the re-entrant SC phase coincides with a reorientation of the U magnetic moments along the \textbf{b}-axis \cite{Levy2005}. In UCoGe, no such anomaly is detected around 14\,T in the magnetization, the reorientation of the moments seems to happen only above 50\,T \cite{Knafo2012}. Non-linear field response of the Shubnikov de Haas frequency is observed above 20\,T \cite{Aoki2011} and a possible field-induced topological Fermi surface (FS) transition, also known as a Lifshitz transition, supported by thermopower \cite{Malone2012}, and magnetoresistivity \cite{Aoki2011,Steven2011} measurements could explain the ``S''-shape of the $H^b_{c2}$ curve at lower field (around 11\,T). Another scenario to explain the observed anomalies of the magnetoresistivity at high fields relies on an unusual polarisability of the U and Co moments detected by polarized neutron experiments \cite{Prokes2010}. However, the corresponding ferro-ferrimagnetic phase transition has never been detected. 

The strong interplay between magnetism and superconductivity is a common feature of the ferromagnetic superconductors. While the magnetism of UGe$_2$ and URhGe is well established and understood, this is not the case for UCoGe where the respective contribution of U and Co is still under debate. There is an urgent need for a detailed knowledge of the magnetism of UCoGe, and precise microscopic studies are now timely owing to the recent progress in single crystal quality of this system. Band structure calculations \cite{Samsel-Czekala2010} and neutron experiments \cite{Prokes2010} have endeavoured to explore the orbital and spin part of the ordered moment, but contradictory results were published. On one hand, theoretical calculations \cite{Samsel-Czekala2010} predict a small uranium moment ($\sim$ 0.1\,$\mu_B$) due to an almost cancellation of substantial orbital and spin moments, and unexpectedly a large cobalt moment (0.2-0.5\,$\mu_B$) either parallel or antiparallel to the U moment. On the other hand comparison of $^{59}$Co NQR and NMR data of YCoGe (a Pauli paramagnet) and UCoGe led to conclude that the ferromagnetism in UCoGe originates predominantly from U-5$f$ electrons at least at low field \cite{Karube2011}. Surprisingly, polarized neutron diffraction experiments \cite{Prokes2010} show that in an applied field of 3\,T, the small ordered moment is essentially carried by the U atoms ($\sim$ 0.1\,$\mu_B$), while at 12\,T a substantial moment ($\sim$ 0.2\,$\mu_B$) antiparallel to the U moment is induced at the Co site and a parallel magnetization is observed in the interstitial regions ($\sim$ 0.3\,$\mu_B$). 

In this letter, we exploit the possibility of X-ray magnetic circular dichroism (XMCD) technique that has emerged the last two decades. This tool which is element and electronic shell specific, allows one to quantitatively estimate the spin and the orbital moments of the absorbing atoms through the use of a set of sum rules \cite{Thole1992,Carra1993}. X-ray absorption near edge structure (XANES) and XMCD at the actinide M$_{4,5}$ edges (3$d\to5f$ transitions) have been demonstrated to be particularly successful for the study of the electronic and magnetic properties of the 5$f$ states in the actinide compounds \cite{Reotier1997,Wilhelm2013}.

\begin{figure}
		\includegraphics[width=0.48\textwidth]{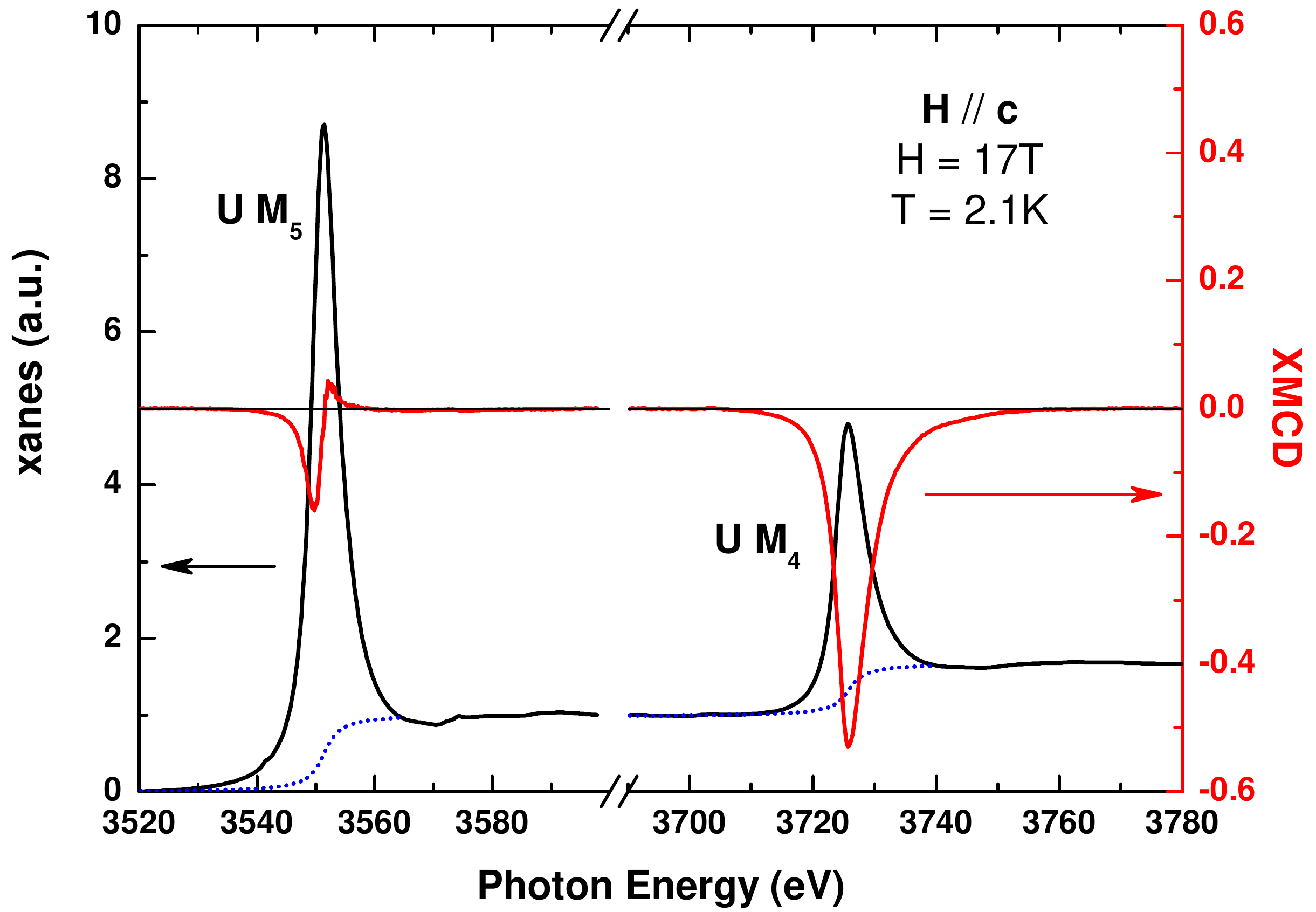}
	\caption{(color online) XANES (black curves, left axis) and XMCD (red curves, right axis) spectra measured at the U-M$_5$ and M$_4$ edges for a UCoGe single crystal in a magnetic field of 17\,T applied along the \textbf{c}-axis at 2.1\,K as a function of the incident photon energy. The integrated areas under the white lines are obtained after subtraction of the continuum modelled by an arctan function (broken lines). The XMCD spectrum has been corrected for self absorption and for incomplete polarization rate.}
	\label{fig1}
\end{figure}

High quality single crystals of UCoGe were grown using the Czochralski method in a tetra-arc furnace followed by annealing under ultra-high vacuum. Details were published elsewhere \cite{Aoki2011a}. The single crystals checked by specific heat and resistivity (residual resistivity RRR = $\rho$(300K)/$\rho$(T$\to$0K), typically of the order of 30) were cut and cleaved before the measurements. The bare shaped crystals of a few mm$^3$, aligned by Laue diffraction and glued on an aluminium support, were mounted on a cryostat (T$\sim$2.1\,K) cold finger inserted in the bore of a superconducting solenoid (H$\leq$17\,T). The magnetic field was parallel either to the easy axis \textbf{c} or to the intermediate axis \textbf{b}. XANES and XMCD experiments were carried out at the ID12 beamline of the European Synchrotron Radiation Facility (ESRF Grenoble) which is dedicated to the polarization-dependent X-ray absorption spectroscopy studies in the photon energy range from 2 to 15\,keV \cite{Rogalev2001} (further experimental details are given in \cite{SupMat}). Element selective magnetization curves were recorded by monitoring the intensity of the XMCD signal at a given photon energy as a function of the applied field.

The isotropic XANES spectrum ($\sigma^+ + \sigma^- + \sigma_0$) of UCoGe at the U-M$_{4,5}$ edges recorded at 2.1\,K in an applied field of 17\,T along the easy axis \textbf{c} was obtained by measuring the absorption spectra at right ($\sigma^+$) and left ($\sigma^-$) polarized X-rays (Fig. \ref{fig1}, black lines). The linear polarized X-ray spectrum $\sigma_0$ was approximated by ($\sigma^+ + \sigma^-)/2$. The spin-orbit sum rule is an useful tool to investigate the nature of the 5$f$ electronic states via the branching ratio \cite{Laan2004}. The branching ratio $B$ for the 3$d_{3/2, 5/2} \to 5f$ transition of U, is experimentally determined as $B = A_{5/2}/(A_{5/2}+A_{3/2})$, where $A_{5/2}$ and $A_{3/2}$ are the integrated areas of the isotropic white lines at the M$_{5,4}$ edges, respectively. $B$ is found to be 0.701(2). This value could be compared with those calculated for different electronic configurations in the intermediate-coupling (IC) approximation which was shown to apply for actinide metals and compounds \cite{Wilhelm2013}. The experimentally determined $B$ is in between those calculated for the 5$f^2$ (U$^{4+}$) and 5$f^3$ (U$^{3+}$) electronic configurations (0.686 and 0.729, respectively). Thus, the 5$f$ electron count is $2<$ n$_e^{5f}<3$. This is in agreement with band structure calculations \cite{Samsel-Czekala2010}, n$_e^{5f}$ is estimated to be 2.84, as well as with core level photoelectron spectroscopy \cite{Fujimori2012} which shows that n$_e^{5f}$ is less but close to 3. These results invalid the occurrence of U$^{4+}$ ions as suggested from neutron form factor analysis \cite{Prokes2010}.

From the branching ratio B, we may determine the expectation value of the angular part of the valence spin-orbit operator as \cite{Laan2004}

\begin{eqnarray}
 	\frac{2<\textbf{l}\cdot\textbf{s}>}{3\mathrm{n}_h^{5f}}=-\frac{5}{2}\left(B-\frac{3}{5}\right)+\Delta
 		\label{eq:SpinOrbitOp_2}
\end{eqnarray}

where n$_h^{5f}$ is the number of holes in the 5$f$ shell, $\Delta$ is a quantity dependent on the electronic configuration. $\Delta$ has been estimated to amount to -0.0106 for the 5$f^{2.84}$ configuration \cite{Laan2004}. Since $<\textbf{l}\cdot\textbf{s}>=\frac{3}{2}\mathrm{n}_e^{5f_{7/2}}-2\mathrm{n}_e^{5f_{5/2}}$, we may evaluate the number of electrons in the individual shells corresponding to j = 7/2 and j = 5/2 if the number of holes in the 5$f$ shell is known (n$_h^{5f} = 14 - $n$_e^{5f}$). Taking the value n$_e^{5f}$ = 2.84 calculated in \cite{Samsel-Czekala2010}, one obtains \mbox{$<\textbf{l}\cdot\textbf{s}> = -4.404$} and the occupation numbers n$_e^{5f_{5/2}}$ = 2.48 and n$_e^{5f_{7/2}}$ = 0.36.

\begin{figure}
		\includegraphics[width=0.48\textwidth]{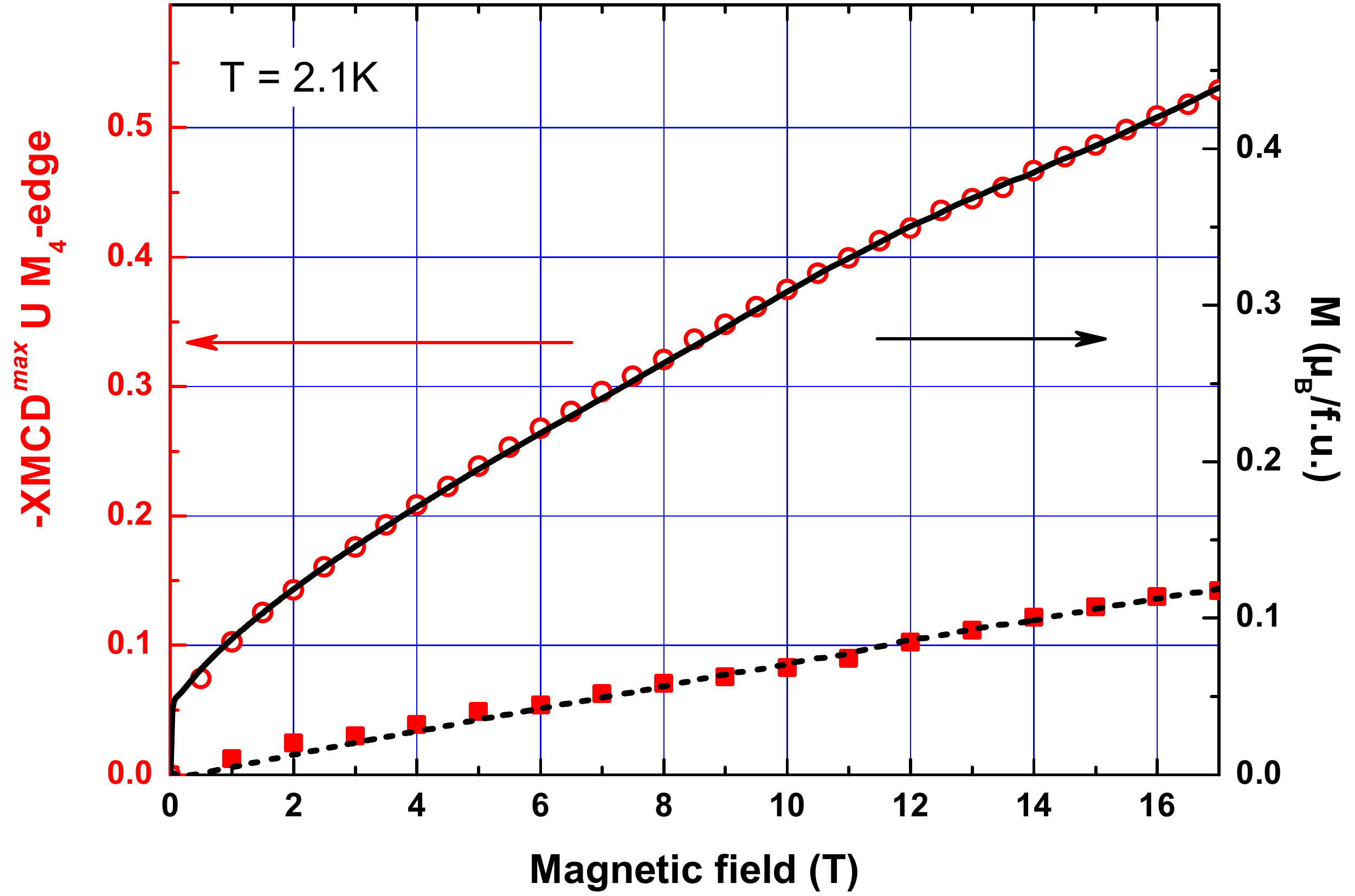}
	\caption{(color online) Uranium magnetization (left axis, red symbols) recorded at the maximum XMCD signal at the U-M$_4$ edge at 2.1\,K with \textbf{H}//\textbf{c} (open circles) and \textbf{H}//\textbf{b} (full squares), together with the macroscopic magnetization curves (right axis, black straight line for \textbf{H}//\textbf{c} and dashed line for \textbf{H}//\textbf{b}), from \cite{Knafo2012}.}
	\label{fig2}
\end{figure}

The XMCD spectrum ($\sigma^+$(E) - $\sigma^-$(E)) recorded at the U-M$_{4,5}$ absorption edges at 2.1\,K under a magnetic field of 17\,T applied along the easy-axis \textbf{c} is shown in Fig. \ref{fig1} (red line). The signal at the M$_4$ edge is large and consists of a negative slightly asymmetric peak. The dichroic signal at the M$_5$ edge is much weaker and presents a ``S'' shape with a negative and a positive peak. This asymmetric shape is known to be sensitive to subtle changes of the electronic structure. It depends strongly on hybridization, crystal field, exchange and Coulomb interactions \cite{Wilhelm2013,Kunes2001}.

\begin{table*}
	\centering
%	\begin{tabular}{|c|c|c|c|c|c|c|c|}
	\begin{tabular*}{1\textwidth}{@{\extracolsep{\fill}} c c c c c c c c}
		\hline \hline
		\multirow{2}{*}{} & \multirow{2}{*}{H(T)} & $\mu_{tot}$ & $\mu_L^U(5f)$ & $\mu_S^U(5f)$ & $\mu_{tot}^U(5f)$ & \multirow{2}{*}{$-\frac{\mu_L^U(5f)}{\mu_S^U(5f)}$} & \multirow{2}{*}{$-\frac{\mu_L^U(5f)}{\mu_S^U(5f)}$ (free ion)}\\
		& & $(\mu_B/atom)$ & $(\mu_B/atom)$ & $(\mu_B/atom)$ & $(\mu_B/atom)$ & & \\
		\hline
%		\multirow{3}{*}{\textbf{H}//\textbf{c}} & 2 ($U^{4+}$) & 17 & 0.44 & 0.749 & -0.203 & 0.545 & 3.68 & 3.36\\
		\multirow{2}{*}{\textbf{H}//\textbf{c}} & 17 & 0.44 & 0.695 & -0.297 & 0.398 & 2.34 & 2.60\\
		 & 1 & 0.09 & 0.135 & -0.059 & 0.076 & 2.29 & 2.60\\
		\hline
		\textbf{H}//\textbf{b} & 17 & 0.12 & 0.165 & -0.080 & 0.085 & 2.06 & 2.60\\
		\hline \hline
	\end{tabular*}
	\caption{Macroscopic moments \cite{Knafo2012}, orbital, spin and total Uranium 5$f$ magnetic moments deduced from the XMCD spectra for different magnetic fields and orientations. The orbital $\mu^U_L$(5$f$) and spin $\mu^U_S$(5$f$) moments were calculated assuming n$^{5f}_e$=2.84 and using the theoretical free ion U$^{3+}$ value (0.62) for the $<T_z>$/$<S_z>$ ratio. The experimental errors for the macroscopic moments are less than 1\% and the error bars in the values of the spin and orbital moments are estimated to be at most of the order of 10 and 5\%, respectively.}
	\label{tab:momentU}
\end{table*} 

The use of the magneto-optical sum rules \cite{SupMat} allows to quantitatively estimate the orbital ($\mu^U_L$(5$f$)) and spin ($\mu^U_S$(5$f$)) moments of the uranium atom. The orbital sum rule provides the $z$-component of the angular momentum $<L_z>$, and the second sum the effective spin polarization $<S_{eff}>$ through the relation $<S_{eff}> = <S_z> + 3<T_z>$ \cite{Carra1993}, where $<S_z>$ is the $z$-component of the ground state expectation value of the spin operator and $<T_z>$ the one of the magnetic dipolar operator.

The $f$-count, n$^{5f}_e$, %estimated from the branching ratio is in-between 2 and 3 and found to be close to 3 from band structure calculations \cite{Samsel-Czekala2010} and core level photoelectron spectroscopy \cite{Fujimori2012}.
 was set to the theoretical band structure value (n$^{5f}_e$ = 2.84) \cite{Samsel-Czekala2010}. %$<T_z>$ is difficult to estimate experimentally and, since UCoGe has a high magnetocrystalline anisotropy, its contribution is expected to be important (as shown for US \cite{Kernavanois2001}) and close to the $<T_z>$ value calculated for the free U$^{3+}$ ion \cite{Kernavanois2001}. In the intermediate coupling approximation $<T_z>$ = 0.62\,$<S_z>$ for the 5$f^3$ (U$^{3+}$) state \cite{Laan1996}.
$<T_z>$ cannot be measured directly but it may be estimated from \textit{ab-initio} band structure calculations or evaluated by combination of XMCD results, polarized neutron diffraction and magnetic Compton scattering data with magnetization measurements \cite{Wilhelm2013,Kernavanois2001}. Since UCoGe has a high magneto-crystalline anisotropy, the $<T_z>/<S_z>$ ratio is expected to be important and close to the value calculated for the free U$^{3+}$ ion in the intermediate coupling scheme (0.62) \cite{Laan1996, Magnani2015}. This was nicely demonstrated in the cases of e.g. US \cite{Kernavanois2001} and UPtAl  \cite{Kucera2002}, which both present a huge magnetic anisotropy. Their $<T_z>/<S_z>$ ratios were found to be 0.83 and 0.65, respectively. The orbital and spin contributions to the uranium magnetic moment are summarized in Table \ref{tab:momentU}.

The choice of n$_e^{5f}$ is not very crucial for the determination of $\mu^U_L$(5$f$). The inaccuracy introduced in the value of the orbital moment is less than a few percents. On the other hand the spin moment comes out with a higher relative error. %The values for $\mu^U_{tot}$(5$f$) and –$\mu^U_L$(5$f$)/$\mu^U_S$(5$f$) assuming n$^{5f}_e$ = 3, i.e. an U$^{3+}$ configuration match better with the macroscopic magnetization and the free ion orbital to spin moment ratio which represents an upper bound.
 The magnetism of U is dominated by the orbital moment, and the spin moment is aligned antiparallel to the orbital component as generally observed in actinide compounds. The comparison of the total uranium moment to the total magnetization (0.44\,$\mu_B$ at 17\,T for \textbf{H}//\textbf{c}, 0.09\,$\mu_B$ at 1\,T for \textbf{H}//\textbf{c}, and 0.12\,$\mu_B$ at 17\,T for \textbf{H}//\textbf{b} \cite{Knafo2012}) indicate that uranium dominates the magnetism of UCoGe. The ratio $-\mu^U_L$(5$f$)/ $\mu^U_S$(5$f$) is about 2.31 at 1\,T and 17\,T for \textbf{H}//\textbf{c} whereas it falls to 2.06 for \textbf{H}//\textbf{b} at 17\,T. This is the first time to our knowledge that a difference in magnetic anisotropy manifests itself through a large change of the orbital to spin moment ratio reflecting the anisotropic character of the 5$f$ electrons. These ratios, which fall below the free ion U$^{3+}$ value (2.60), indicate a significant delocalization of the 5$f$-electron states due to the hybridization of the U-5$f$ electrons with the conduction band and Co-3$d$ electrons \cite{Lander1991}. 

Figure \ref{fig2} demonstrates that the magnetization curve recorded at the maximum XMCD signal at the M$_4$ edge matches well with the macroscopic high-field magnetization \cite{Knafo2012} (black lines) in the whole range of applied fields when \textbf{H}//\textbf{b} (red full squares) and \textbf{H}//\textbf{c} (red open circles). Note that in the later configuration, in this field range (0-17\,T), the slope of M(H) decreases with increasing field. When \textbf{H}//\textbf{b}, the U-magnetization is linear up to 17\,T and shows no anomaly at H$\sim$14\,T, field where $H^b_{c2}$ displays the ``S''-shape. This overall behavior suggests that the Co/Ge and the conduction electron contributions to the macroscopic magnetization should have the same H dependence as the U-5$f$ moment.

The total magnetization data gives the total moment $\mu_{tot}$ = $\mu^U_{tot}$(5$f$) + $\mu^{cond}$ + $\mu^{Co}$ + $\mu^{Ge}$, where $\mu^{cond}$ is the contribution, usually small, from the uranium 6$d$7$s$ conduction band. A rough estimate is that it is about $-10$\% of the total magnetization \cite{Wilhelm2013,Rossat-Mignod1984}. The electron conduction contribution is negative because these electrons are polarized by the spin contribution of the uranium which is antiparallel to the U moment, dominated by the orbital moment. It is safe to assume that the contribution from the Ge atoms can be ignored \cite{Kernavanois2001a}. Thus for 17\,T applied along the easy-axis \textbf{c}, $\mu^{Co}=\mu_{tot}-\mu^U_{tot}(5f)+0.1\,\mu_{tot}\sim +0.09$\,$\mu_B$ (the error in the value of $\mu^{Co}$ is estimated to be less than 0.02\,$\mu_B$). This is to be compared to the total magnetization of 0.44\,$\mu_{B}$. The observed small Co moment  parallel to the bulk (or 5$f$) magnetization is expected to originate mainly from the polarization of the Co-3$d$ band strongly hybridized to the U-5$f$ band \cite{Samsel-Czekala2010}. The parallel orientation of the U-5$f$ and Co-3$d$ moments is not surprising if one considers that according to the mechanism proposed by Brooks \textit{et al.} \cite{Brooks1989}, the 3$d$-spin moments of the Co atoms are coupled antiferromagnetically to the 6$d$-spin moments of the U atoms. Due to the positive intra-atomic Hund's exchange coupling the U-5$f$ spin moments are in turn coupled parallel to U-6$d$ spin moment and therefore antiparallel to the Co 3$d$-spin. The parallel alignment of the uranium and cobalt-3$d$ magnetic moments may also be concluded from magnetic Compton scattering experiments \cite{Butchers2013} and from XMCD measurements at the Co-L$_{2,3}$ edges \cite{Butchers2013}. Although the XMCD spectra similar to those of UCoAl \cite{Takeda2013} are difficult to interpret quantitatively owing to the overlap of the L$_3$ edge of cobalt (778.1\,eV) with the N$_4$ edge of U (778.4\,eV), and possible surface effects, a Co moment of about 0.05\,$\mu_B$ at 6\,T could be estimated \cite{Butchers2013}, yielding to 0.096\,$\mu_B$ at 17\,T (see Fig. \ref{fig4}), within error bars of our estimate (0.09\,$\mu_B$).

\begin{figure}
		\includegraphics[width=0.48\textwidth]{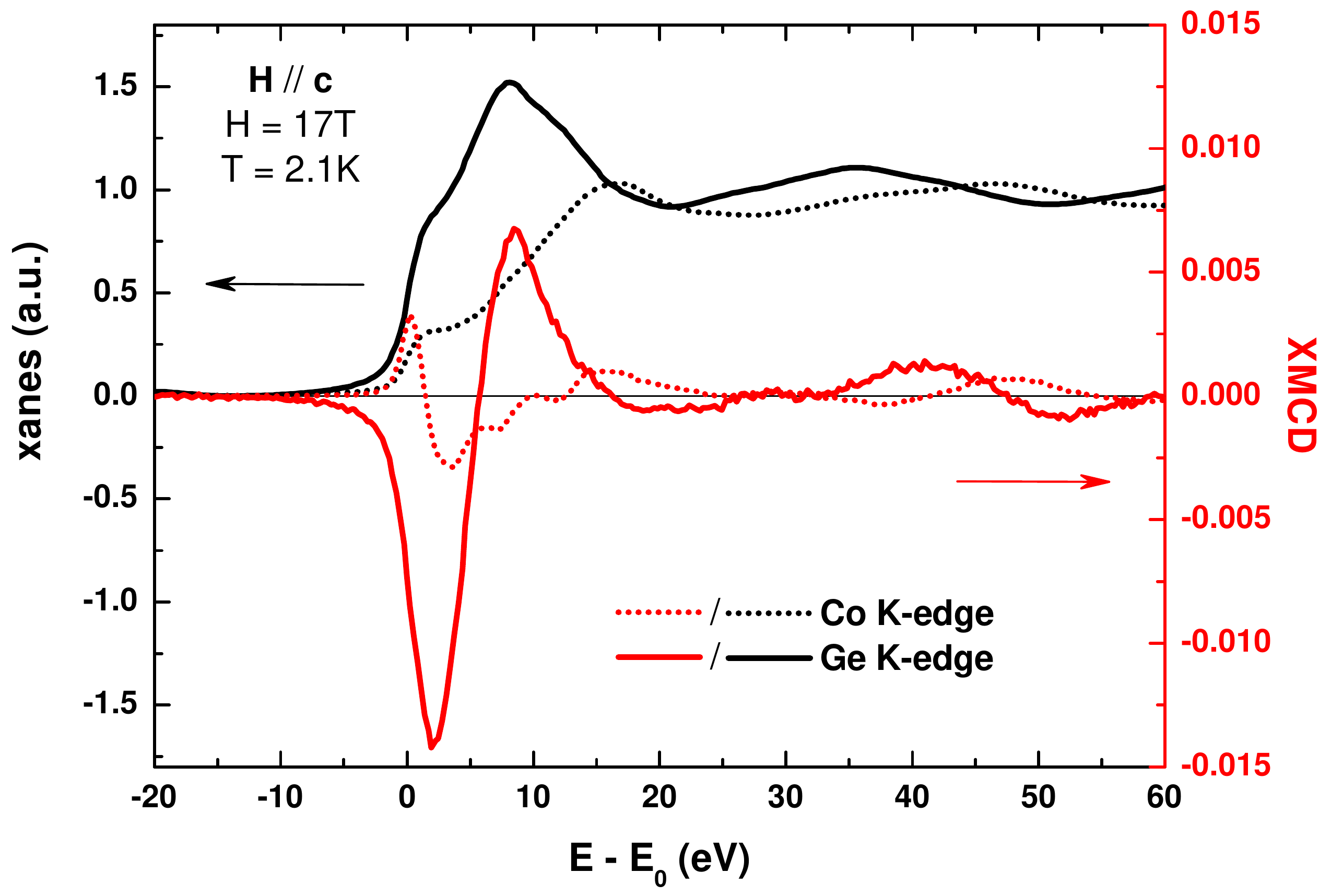}
	\caption{(color online) XANES (left axis, black curves) and XMCD (right axis, red curves) spectra recorded at the K edges of Co (dotted lines) and Ge (straight lines) at 2.1\,K and 17\,T with \textbf{H}//\textbf{c}.}
	\label{fig3}
\end{figure}

\begin{figure}
		\includegraphics[width=0.48\textwidth]{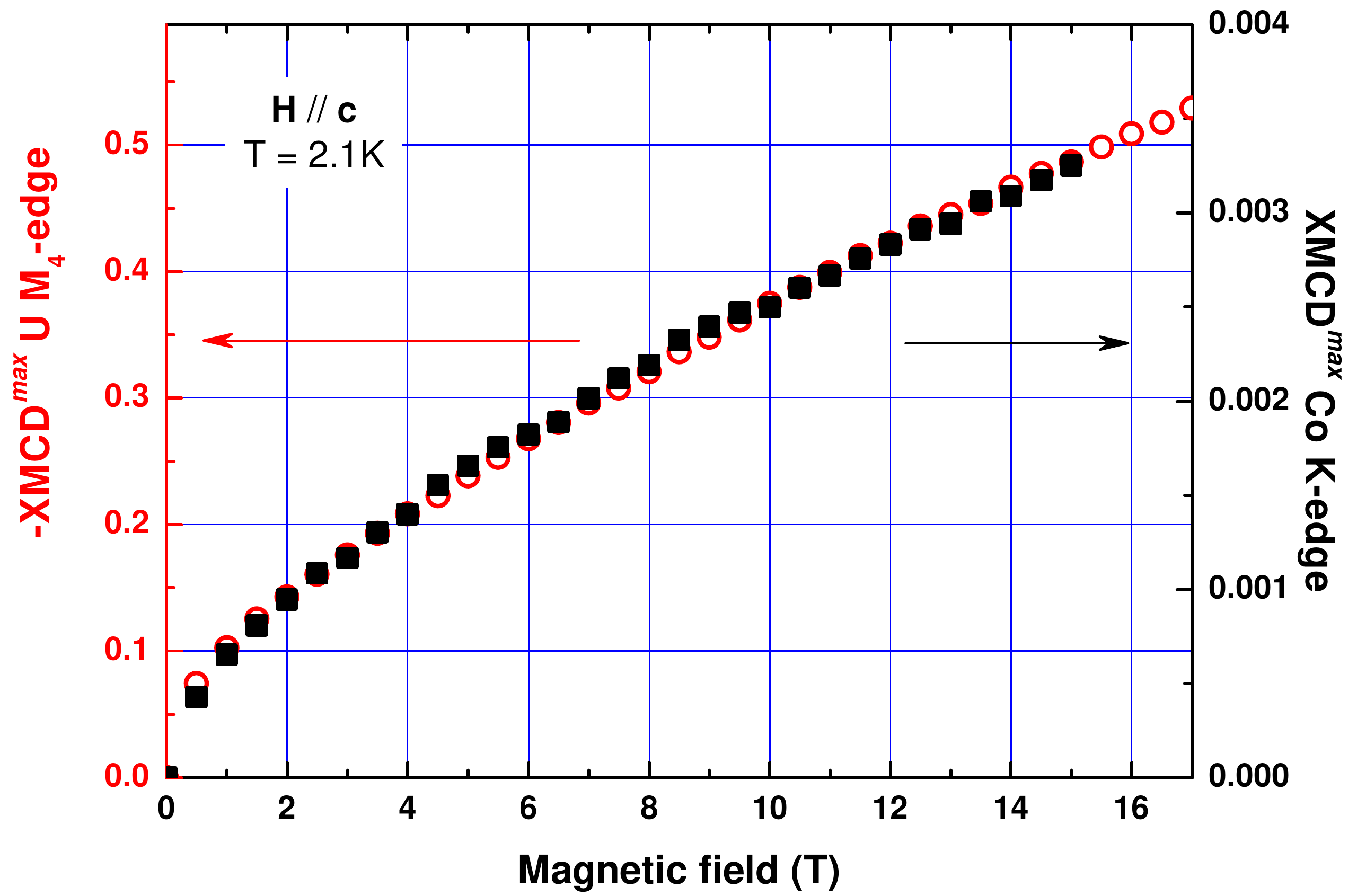}
	\caption{(color online) Element specific magnetization recorded at the maximum XMCD signal at the Co-K absorption edge (right axis, black squares). Comparison with the U-magnetization measured at the maximum XMCD signal at the U-M$_4$ edge at 2.1\,K with \textbf{H}//\textbf{c} (left axis, red open circles).}
	\label{fig4}
\end{figure}

Normalized XANES and XMCD spectra recorded at the K edges of Co (and Ge) at 2.1\,K and 17\,T applied along the \textbf{c}-axis are presented in Fig. \ref{fig3}. The XMCD signal at the K edges is weak and more intricate to interpret, because it is due only to the orbital polarization of the 4$p$ states that is induced either via intra-atomic spin-orbit coupling when there is a sizeable local spin moment or by hybridization of the 4$p$ states with spin-orbit split 5$f$ states of uranium ions \cite{Solovyev1995,Rueff1998} (the contribution of the electric quadrupole transition is extremely small and is not considered here) \cite{Igarashi1996}. Positive and negative peaks show up at the Co/Ge-K edges. The oscillations observed above the K edges ($>$20\,eV) arise from magnetic EXAFS (Extended X-ray Fine Structure) related to the magnetic local surroundings of the Co/Ge atoms. The presence of a small 4$p$-orbital Co moment is inferred from the XMCD data at the Co-K edge. The integration (up to 20\,eV above the edge) of the sharp positive peak and of the broad negative peak results in a reduced negative signal, i.e, a much smaller positive orbital magnetic moment of the Co-4$p$ state compared to the one of the Ge-4$p$ state. Regarding Ge, one has to say that its 3$d$ band, which is fully filled and lies below the Fermi energy, does not carry a moment. The only moment on the Ge site is the one induced by the neighboring atoms U or/and Co. From extensive studies of rare-earth (R) - 3$d$ transition metal (T) intermetallics, Boada \textit{et al.} \cite{Boada2010} concluded that both R and T sublattices contribute to the XMCD at the T-K edge in an additive way. Thus, it was tempting to compare the Co-K edge XMCD of UCoGe with the one of YCoGe where Y is non-magnetic \cite{SupMat}. In YCoGe, the sharp positive peak observed in UCoGe at the Co-K edge in absent and the broad negative contribution is strongly reduced. According to the argument of Boada \textit{et al.} \cite{Boada2010}, the positive peak in UCoGe could tentatively be attributed to the U contribution only, but theoretical and further experimental works are needed to support this conclusion.

The field dependence of the maximum signal at the Co-K edge is depicted in Fig. \ref{fig4} where it is compared to the U-magnetization curve. The Co-4$p$ magnetization follows nicely the U-magnetization. This behavior is in contrast with the polarized neutron diffraction data \cite{Prokes2010} which showed that $|\mu^{Co}/\mu^{U}|$ varies from 0.3 to 0.84 between 3 and 12\,T at 100\,mK (in contrast to XMCD Co-K edge data, the neutron measurements probe the Co-3$d$ moment. However, both Co-4$p$ and Co-3$d$ moments are induced by the U-moments. Their field dependences are thus similar).

%The similar field dependences of the local U/Co magnetizations and the macroscopic magnetization indicate that U induces the Co moment. 
Our data also allow to rule out that the magnetoresistivity anomaly observed for \textbf{H}//\textbf{c} at $\sim$ 8\,T (at 3\,K) \cite{Aoki2011,Prokes2010,Bay2014} may be due to a ferromagnetic-ferrimagnetic transition as suggested by neutron diffraction \cite{Prokes2010}. These anomalies may thus be linked, as suggested before, to a change of the FS under magnetic field \cite{Aoki2011}.

In conclusion, XMCD studies of UCoGe reveal that the $f$-count in the uranium 5$f$ band is close to 3 and that the U-5$f$ electrons drive the magnetism and the superconductivity of this ferromagnetic superconductor. The observed parallel weak Co-3$d$ moment is induced by hybridization with the U-5$f$ states. Its magnitude is at most 20\% of the bulk moment. The reduced (and anisotropic) value of the orbital to spin U moments ratio indicates a significant (anisotropic) U-5$f$/Co-3$d$ hybridization. Contrary to polarized neutron diffraction studies, our XMCD data neither confirm the antiparallel coupling of the U and Co moment nor support that the Co moment compares to the U moment. They provide a strong data basis for more accurate band structure calculations and for further efforts on polarized neutron experiments.
\newline

%\section*{Acknowledgement}
We gratefully thank W. Knafo for providing his magnetization data, as well as N. Sato and K. Ishida and for providing us unpublished susceptibility data on YCoGe. This work has been supported by the French ANR grant "SINUS" and the ERC grant ``NewHeavyFermion''. One of us (D.A.) acknowledges support from KAKENKI, ICC-IMR, and REIMEI programs.

%\begin{thebibliography}{99}

%merlin.mbs apsrev4-1.bst 2010-07-25 4.21a (PWD, AO, DPC) hacked
%Control: key (0)
%Control: author (8) initials jnrlst
%Control: editor formatted (1) identically to author
%Control: production of article title (-1) disabled
%Control: page (0) single
%Control: year (1) truncated
%Control: production of eprint (0) enabled
%

%\bibliography{biblio,/Users/jeanpascalbrison/biblioLatex/bibliography}
%\bibliography{biblio}

\newpage
\part{Supplementary Material}
\section{1. Experimental details}
For the experiments at the U-M$_4$ (3.728\,keV) and M$_5$ (3.552\,keV) edges, the source was the helical undulator Helios II which provides high flux of circular polarized X-ray photons with a polarization rate close to 1. After monochromatization with a double crystal Si(111), the polarization was reduced to $\sim$ 0.45 at the M$_4$ edge and $\sim$ 0.35 at the M$_5$ edge. The XANES at both the U-M$_{4,5}$ edges and at the Co/Ge-K edges (7.709 and 11.103\,keV respectively) were recorded using the total fluorescence yield detection mode in the backscattering geometry for parallel $\sigma^+$(E) and antiparallel $\sigma^-$(E) alignments of the photon helicity with respect to the external field applied along the beam direction. 
The XANES spectra for right and left circular polarized X-ray beams were corrected assuming practically infinite thickness of the samples, but taking into account the various background contributions (fluorescence of subshells and matrix as well as coherent and incoherent scattering), the angle of incidence of the X-ray beam, the solid angle of the detector, the incomplete polarization rates of incident X-ray photons \cite{Goulon1982,Troger1992,Pfalzer1999}. The U edge jump intensity ratio M$_5$:M$_4$ was normalized to 1:2/3 according to the statistical edge jump ratio (defined as the ratio between the occupation numbers for the two spin-orbit split core levels j = 3/2 and j = 5/2) \cite{Berger}. Regarding the Co/Ge-edges, the spectra were also corrected for self-absorption after normalization of the edge jump to unity.

\section{2. Sum rules}
\label{AppB}
The orbital sum rule links the integrated dichroic signals over the two M$_{4,5}$ edges to the ground state expectation value of the $z$-component of the angular momentum L acting on the 5$f$ shell which receives the photoelectron in the final state. For the 3$d\to5f$ transition the orbital sum rule writes \cite{Thole1992}: 

\begin{align*}
 	<L_z>=\frac{3\mathrm{n}_h^{5f}\bigintss_{M_4+M_5}\Delta\sigma(E)\,\mathrm{d}E}{\bigintss_{M_4+M_5}(\sigma^+(E) +\sigma^-(E) + \sigma_0(E))\,\mathrm{d}E}
\end{align*}

$\Delta\sigma = \sigma^+ -\sigma^-$ corresponds to the dichroism, E is the photon energy. The second sum rule correlated a linear combination of the partial dichroism signals at the M$_{4,5}$ edges with the effective spin polarization $<S_{eff}>$ which is related to the spin operator through the relation \cite{Carra1993} 

\begin{align*}
 	<S_{eff}> &= <S_z> + 3<T_z>\\
 	 &= \frac{3}{4}\mathrm{n}_h^{5f}\frac{2\bigintss_{M_5}\Delta\sigma(E)\,\mathrm{d}E-3\bigintss_{M_4}\Delta\sigma(E)\,\mathrm{d}E}{\bigintss_{M_4+M_5}(\sigma^+(E) +\sigma^-(E) + \sigma_0(E))\,\mathrm{d}E}
\end{align*}
 
$<S_z>$ is the $z$-component of the ground state average value of the spin operator and $<T_z>$ the one of the magnetic dipolar operator. $<T_z>$ is related to the anisotropy of the local magnetic field produced by the spin when the valence cloud is distorted either by spin orbit and/or crystal field interactions. The orbital $\mu^U_L$(5$f$) and spin $\mu^U_S$(5$f$) components of the total uranium moment
\[
\mu^U_{tot}(5f) = - (<L_z> + 2<S_z>)\,\mu_B
\]
 can be obtained from XMCD spectra if the 5$f$ occupation number and $<T_z>$ are known.

\section{3. XMCD at the Co-edge of YCoGe}

\begin{figure}
		\includegraphics[width=0.75\textwidth]{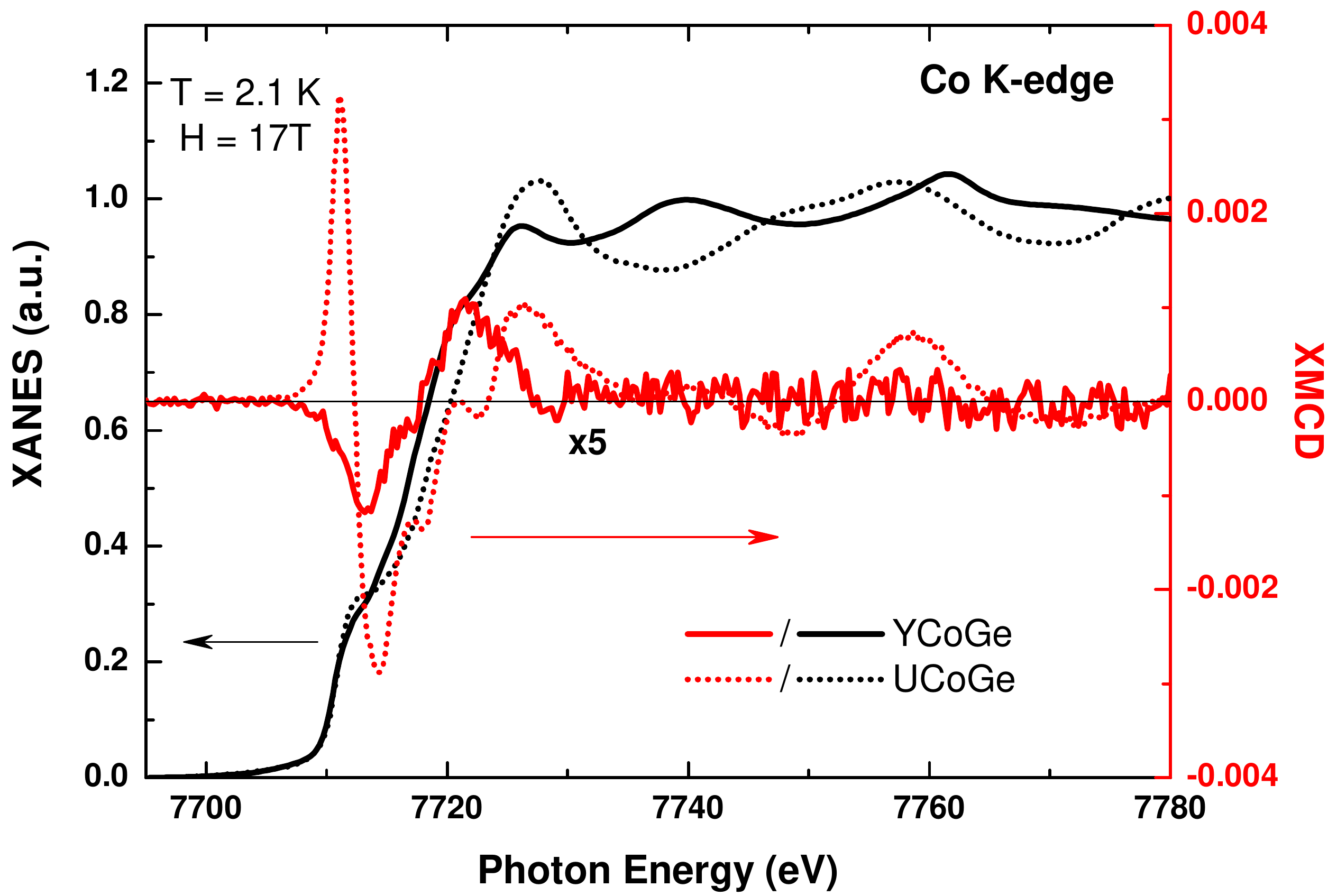}
	\caption{(color online) XANES (left axis, in black) and XMCD (right axis, in red) spectra of an UCoGe single crystal (\textbf{H}//\textbf{c}, dotted lines) and of an YCoGe polycrystal (straight lines) recorded at the K-edge of Co at 2.1\,K and 17\,T.}
	\label{figS1}
\end{figure}

The YCoGe polycrystalline sample was prepared by arc melting of the elements and characterized by X-ray diffraction. Its structure, similar to the UCoGe one (TiNiSi-type), differs only in the alignment of Co-Ge \cite{Karube2011}. It was shown that YCoGe exhibits a typical metallic behavior without magnetic and superconducting anomalies down to 0.3\,K \cite{Karube2011}. In Figure S\ref{figS1}, the XANES and XMCD spectra of YCoGe recorded at 2.1\,K and 17\,T at the Co-edge are presented and compared to those of UCoGe.

The dichroic signal at the Co-K edge of YCoGe exhibits a broad negative structure centered at about 4.5\,eV above the absorption edge. It resembles to the one observed in hcp Co metal \cite{Rueff1998}, but with a much smaller intensity (about a factor 10), as expected since YCoGe is an almost temperature independent paramagnet ($\mu \sim 10^{-4}\,\mu_B/$mole at 17\,T) \cite{IshidaPrivate}, whereas Co(hcp) is a ferromagnet ($\mu=1.67\,\mu_B/$atom). On the other hand, the XMCD spectrum obtained from UCoGe strongly contrasts with that of YCoGe, where Y is non-magnetic. In UCoGe, the dichroic signal consists of a sharp positive peak at $\sim$2.5\,eV above the absorption edge and a large asymmetric negative structure with a maximum depth at $\sim$10\,eV. The oscillations observed ($\approx$20\,eV) above the K-edge may be ascribed to magnetic EXAFS. The integration (up to 20\,eV above the edge) of the sharp positive peak and the broad negative structure results in a negative signal, i.e. a positive (parallel to \textbf{H}) orbital Co-4$p$ moment significantly larger than the one of YCoGe as expected. In UCoGe, both U and Co contribute to the observed dichroic signal. Following Boada \textit{et al.} \cite{Boada2010}, it is tempting to attribute the sharp positive peak observed in UCoGe and absent in YCoGe to the U contribution alone, but it is more intricate to disentangle the respective contributions to the negative structure. Further theoretical as well as experimental (e.g. XMCD of U$_{1-x}$Y$_x$CoGe solid solutions) works are needed to sort out the different contributions to the dichroic signal in UCoGe.

%merlin.mbs apsrev4-1.bst 2010-07-25 4.21a (PWD, AO, DPC) hacked
%Control: key (0)
%Control: author (8) initials jnrlst
%Control: editor formatted (1) identically to author
%Control: production of article title (-1) disabled
%Control: page (0) single
%Control: year (1) truncated
%Control: production of eprint (0) enabled
%

%\end{thebibliography}
\end{document}